\documentclass[showpacs,aps,graphicx,twocolumn]{revtex4}%and
\usepackage{graphicx}

\begin{document}

\title{High efficient multipartite entanglement purification using hyperentanglement}

\author{Lan Zhou,$^{1}$ Pei-Shun Yan,$^{2}$ Wei Zhong,$^{2}$ Yu-Bo Sheng$^{2}$\footnote{shengyb@njupt.edu.cn} }
\address{$^1$ School of Science, Nanjing University of Posts and Telecommunications, Nanjing,
210003, China\\
$^2$Institute of Quantum Information and Technology, Nanjing University of Posts and Telecommunications, Nanjing, 210003,  China\\}

\date{\today }
\begin{abstract}
Multipartite entanglement plays an important role in controlled quantum teleportation, quantum secret sharing, quantum metrology and some other important quantum information branches. However, the maximally multipartite entangled state will degrade into the mixed state because of the noise. We present an efficient multipartite entanglement purification protocol (EPP) which can distill the high quality entangled states from low quality entangled states for $N$-photon systems in a Greenberger-Horne-Zeilinger (GHZ) state in only linear optics.  After performing the protocol, the spatial-mode entanglement is used to purify the polarization entanglement and one pair of high quality polarization entangled state will be obtained. This EPP has several advantages. Firstly, with the same purification success probability, this EPP only requires one pair of multipartite GHZ state, while existing EPPs usually require two pairs of multipartite GHZ state. Secondly, if consider the practical transmission and detector efficiency, this EPP may be extremely useful for the ratio of  purification efficiency is increased rapidly with both the number of photons and the transmission distance. Thirdly, this protocol requires linear optics and does not add additional measurement operations, so that it is feasible for experiment.  All these advantages will make this protocol have potential application for future quantum information processing.

\end{abstract}
\pacs{ 03.67.Lx} \maketitle
\section{Introduction}
Entanglement plays an important role in quantum communication and computation. Quantum teleportation \cite{teleportation}, quantum key distribution\cite{QKD}, dense coding\cite{densecoding}, quantum secure direct communication\cite{QSDC1,QSDC2,QSDC3}, distributed quantum computing \cite{computation}, distributed secure quantum machine learning \cite{DSQML}, and other important branches all require the parties to share the entanglement. Besides the bipartite entanglement, multipartite entanglement also plays an important role in controlled quantum teleportation \cite{cteleportation1,cteleportation2}, quantum secret sharing\cite{QSS1,QSS2,QSS3}, quantum state sharing\cite{QSTS1,QSTS2,QSTS3}, quantum metrology\cite{metrology1,metrology2}, and so on.   Recently, the multipartite entangled state named Greenberger-Horne-Zeilinger (GHZ) states was been used in some important quantum communication experiment, such as long-distance measurement-device-independent multiparty quantum communication \cite{chenzb}, equitable multiparty quantum communication without a trusted third party \cite{jeong1} and quantum teleportation of shared quantum secret \cite{jeong2}. The GHZ state also have been realized  with  superconducting
system \cite{superconduct1,superconduct2},  trapped ions \cite{ion}, and  photonic system \cite{photon}. The detection of the multipartite entanglement structure has also been reported \cite{detection}.

In a practical application, the quantum system should inevitably interact with its environment, and the environment noise will degrade the entanglement. In general, the decoherence will make the maximally entangled state become a mixed state. The degraded entanglement will decrease the efficiency of the quantum communication and it also will make the quantum communication become insecure \cite{repeater}. Entanglement purification is a powerful tool to distill the high quality entangled states from the low quality entangled states \cite{purification1,purification2,purification3,purification4,purification5,addpurification1,addpurification3,experiment2,purification6,purification7,purification8,purification9,
purification11,purification12,purification13,
purification14,purification15,purification16,purification17,purification18,purification19,purification20,purification21,purification22,purification23,
purification24,purification25,purification26,addpurification2,purification27,addpurification4,shengprl}. Entanglement purification has been widely discussed since Bennett et al. proposed the first entanglement purification protocol (EPP) \cite{purification1}. EPPs for bipartite system were proposed in photonic system \cite{purification3,purification4,purification5,purification7,purification8,purification9}, elections\cite{addpurification1}, quantum dots\cite{purification11,purification12}, atoms \cite{experiment2,purification13} and so on. For example, in 2001, Pan et al. proposed the polarization EPP with only feasible linear optics elements \cite{purification3}. In 2008, the EPP with spontaneous parametric down conversion (SPDC) source based on cross-Kerr nonlinearity was proposed \cite{purification7}. In this EPP, the purified high quality entangled state can be remained for further application and the remained entangled states can also be repeated to perform the purification to obtain the higher entangled states.
 The experiments of entanglement purification in optical system were also reported \cite{purification4}. In 2017, Chen et al. realized the nested entanglement purification for quantum repeaters. In this experiment, the entanglement purification and entanglement swapping can be realized simultaneously \cite{purification17}. On the other hand, the double-pair noise components from the SPDC source can be eliminated automatically.  This work was extended to the multi-copy cases \cite{addpurification2}. The optimal entanglement purification was also investigated \cite{purification26}. Recently, the first high efficient and long-distance entanglement purification using hyperentanglement was demonstrated \cite{shengprl}. The hyperentanglement was first distributed to 11 km and the spatial entanglement was used to purification polarization entanglement. The authors also demonstrated its powerful application in entanglement-based QKD.

For multipartite system, Murao et al. described the first EPP with controlled-not (CNOT) gate \cite{multipurification1}. In 2003, Dur et al. described the EPP for Graph state \cite{multipurification2}. In 2007, this protocol was extended to high-dimension multipartite system with the generalized CNOT gate \cite{multipurification3}. In 2008, the  multipartite EPP for polarization entangled states with cross-Kerr nonlinearity was proposed \cite{multipurification4}. In 2011, Deng proposed the multipartite EPP using entanglement link from subspace \cite{multipurification5}. In his protocol, the discussed items in conventional EPPs still have entanglement in a subspace and they can be reused with entanglement link.  There are another kind of EPP for multipartite entanglement system, named deterministic EPP \cite{multipurification6,multipurification7}. In these EPPs, they exploit the hyperentanglement to perform the purification. Such EPPs are based on the condition that the spatial mode entanglement is robust  and it does not suffer from the noise. Therefore, the spatial mode entanglement or the frequency entanglement can be completely transformed to the polarization entanglement. In conventional EPPs, they all require two copies of low quality entangled states. After performing the CNOT or similar operations, one pair of high quality entangled state is remained, if the purification is successful. On the other hand, if the purification is a failure, both pairs should be discarded.

In this paper, we will describe an efficient EPP for multipartite polarization
entangled systems in a GHZ state, inspired the idea of Ref.\cite{shengprl} . Different from existing EPPs for multipartite system, this protocol only requires one pair of hyperentangled state. By performing the CNOT gate between two degrees of freedom, the spatial entanglement is consumed. Therefore, if the protocol is successful, one can obtain a high quality polarization entangled state. This EPP is based on linear optics and it is also feasible for current experiment condition.

This protocol is organized as follows. In Sec.II, we describe this EPP for bit-flip error. In Sec.III, we describe this protocol for phase-flip error. In Sec.IV, we extend this EPP to a general case for arbitrary N-photon GHZ state. In Sec.V, we  present a discussion. Finally, in Sec. VI, we will provide a conclusion.

\section{Multipartite entanglement purification for bit-flip error}
In this section, we describe this EPP with a simple example. The three-photon GHZ states can be written as follows.
\begin{eqnarray}
|\Phi_{0}^{\pm}\rangle_{ABC}=\frac{1}{\sqrt{2}}(|H\rangle_{A}|H\rangle_{B}|H\rangle_{C}\pm|V\rangle_{A}|V\rangle_{B}|V\rangle_{C}),\nonumber\\
|\Phi_{1}^{\pm}\rangle_{ABC}=\frac{1}{\sqrt{2}}(|H\rangle_{A}|H\rangle_{B}|V\rangle_{C}\pm|V\rangle_{A}|V\rangle_{B}|H\rangle_{C}),\nonumber\\
|\Phi_{2}^{\pm}\rangle_{ABC}=\frac{1}{\sqrt{2}}(|H\rangle_{A}|V\rangle_{B}|H\rangle_{C}\pm|V\rangle_{A}|H\rangle_{B}|V\rangle_{C}),\nonumber\\
|\Phi_{3}^{\pm}\rangle_{ABC}=\frac{1}{\sqrt{2}}(|V\rangle_{A}|H\rangle_{B}|H\rangle_{C}\pm|H\rangle_{A}|V\rangle_{B}|V\rangle_{C}).\label{polarization}
\end{eqnarray}
Here $|H\rangle$ denotes the horizonal polarization and $|V\rangle$ denotes the vertical polarization of the photon, respectively.
\begin{figure}[!h]%[tpb]
\begin{center}
\includegraphics[width=8cm,angle=0]{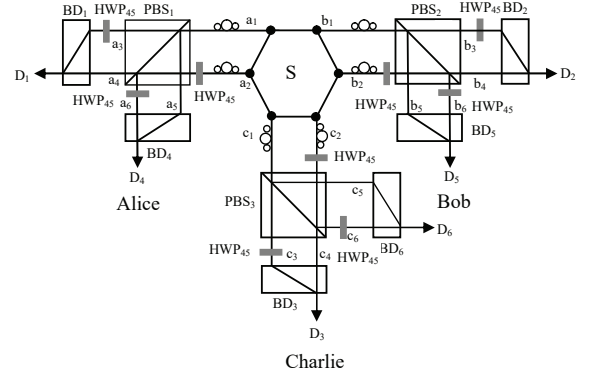}
\caption{Schematic drawing showing the principle of entanglement purification for bit-flip error. HWP$_{45}$ is the half-wave plate setting as 45$^{\circ}$. The PBS is the polarization beam splitter which can transmit the $|H\rangle$ polarization and reflect the $|V\rangle$ polarization photon. The polarizing beam displacers (BD) can couple $|H\rangle$ and $|V\rangle$ polarization
components from different spatial modes.}
\end{center}
\end{figure}
The spatial mode GHZ states can be written as follows.
\begin{eqnarray}
|\phi_{0}^{\pm}\rangle_{ABC}=\frac{1}{\sqrt{2}}(|a_{1}\rangle_{A}|b_{1}\rangle_{B}|c_{1}\rangle_{C}\pm|a_{2}\rangle_{A}|b_{2}\rangle_{B}|c_{2}\rangle_{C}),\nonumber\\
|\phi_{1}^{\pm}\rangle_{ABC}=\frac{1}{\sqrt{2}}(|a_{1}\rangle_{A}|b_{1}\rangle_{B}|c_{2}\rangle_{C}\pm|a_{2}\rangle_{A}|b_{2}\rangle_{B}|c_{1}\rangle_{C}),\nonumber\\
|\phi_{2}^{\pm}\rangle_{ABC}=\frac{1}{\sqrt{2}}(|a_{1}\rangle_{A}|b_{2}\rangle_{B}|c_{1}\rangle_{C}\pm|a_{2}\rangle_{A}|b_{1}\rangle_{B}|c_{2}\rangle_{C}),\nonumber\\
|\phi_{3}^{\pm}\rangle_{ABC}=\frac{1}{\sqrt{2}}(|a_{2}\rangle_{A}|b_{1}\rangle_{B}|c_{1}\rangle_{C}\pm|a_{1}\rangle_{A}|b_{2}\rangle_{B}|c_{2}\rangle_{C}).\label{spatial}
\end{eqnarray}
Here a$_{1}$, a$_{2}$, b$_{1}$, b$_{2}$, c$_{1}$ and c$_{2}$ are the spatial modes as shown in Fig. 1. The PBS is the polarization beam splitter which can transmit the $|H\rangle$ polarization and reflect the $|V\rangle$ polarization photon. The polarizing beam displacers (BD) can couple $|H\rangle$ and $|V\rangle$ polarization
components from different spatial modes. HWP$_{45}$ is the half-wave plate setting as 45$^{\circ}$. It can convert $|H\rangle$  polarization to $|V\rangle$  and $|V\rangle$  to $|H\rangle$ , respectively.
The entanglement source $S$ emits a pair of hyperentangled GHZ state of the form
\begin{eqnarray}
|\Psi\rangle_{ABC}=|\Phi_{0}^{+}\rangle_{ABC}\otimes|\phi_{0}^{+}\rangle_{ABC}.
\end{eqnarray}
The hyperentangled GHZ state is distributed to Alice, Bob and Charlie, respectively. During distribution, if a bit-flip error occurs on both the polarization and spatial modes entangled state, the original state $|\Psi\rangle_{ABC}$ will become a mixed state as
\begin{eqnarray}
\rho_{ABC}=\rho^{P}_{ABC}\otimes\rho^{S}_{ABC}.\label{mixed1}
\end{eqnarray}
Here $\rho^{P}_{ABC}$ and $\rho^{S}_{ABC}$ are the mixed state in polarization and spatial modes. They can be written as
\begin{eqnarray}
\rho^{P}_{ABC}=F_{1}|\Phi_{0}^{+}\rangle_{ABC}\langle\Phi_{0}^{+}|+(1-F_{_{1}})|\Phi_{1}^{+}\rangle_{ABC}\langle\Phi_{1}^{+}|,\label{polarization1}
\end{eqnarray}
and
\begin{eqnarray}
\rho^{S}_{ABC}=F_{2}|\phi_{0}^{+}\rangle_{ABC}\langle\phi_{0}^{+}|+(1-F_{_{2}})|\phi_{1}^{+}\rangle_{ABC}\langle\phi_{1}^{+}|.\label{spatial1}
\end{eqnarray}
From Eq.(\ref{mixed1}), the mixed state $\rho_{ABC}$ can be described as follows. With the probability of $F_{1}\otimes F_{2}$, it is in the state $|\Phi_{0}^{+}\rangle_{ABC}\otimes|\phi_{0}^{+}\rangle_{ABC}$. With the probability of $(1-F_{1})(1-F_{2})$, it is in the state $|\Phi_{1}^{+}\rangle_{ABC}\otimes|\phi_{1}^{+}\rangle_{ABC}$. With the probability of $F_{1}(1-F_{2})$ and $(1-F_{_{1}})F_{2}$, they are in the states $|\Phi_{0}^{+}\rangle_{ABC}\otimes|\phi_{1}^{+}\rangle_{ABC}$ and $|\Phi_{1}^{+}\rangle_{ABC}\otimes|\phi_{0}^{+}\rangle_{ABC}$, respectively. The first case $|\Phi_{0}^{+}\rangle_{ABC}\otimes|\phi_{0}^{+}\rangle_{ABC}$ can be described as
\begin{eqnarray}
&&|\Phi_{0}^{+}\rangle_{ABC}\otimes|\phi_{0}^{+}\rangle_{ABC}\nonumber\\
&=&\frac{1}{\sqrt{2}}(|H\rangle_{A}|H\rangle_{B}|H\rangle_{C}+|V\rangle_{A}|V\rangle_{B}|V\rangle_{C})\nonumber\\
&\otimes&\frac{1}{\sqrt{2}}(|a_{1}\rangle_{A}|b_{1}\rangle_{B}|c_{1}\rangle_{C}+|a_{2}\rangle_{A}|b_{2}\rangle_{B}|c_{2}\rangle_{C})\nonumber\\
&=&\frac{1}{2}(|H\rangle_{a_{1}}|H\rangle_{b_{1}}|H\rangle_{c_{1}}+|H\rangle_{a_{2}}|H\rangle_{b_{2}}|H\rangle_{c_{2}}\nonumber\\
&+&|V\rangle_{a_{1}}|V\rangle_{b_{1}}|V\rangle_{c_{1}}+|V\rangle_{a_{2}}|V\rangle_{b_{2}}|V\rangle_{c_{2}})\nonumber\\
&\rightarrow&\frac{1}{2}(|V\rangle_{a_{3}}|V\rangle_{b_{3}}|V\rangle_{c_{3}}+|H\rangle_{a_{6}}|H\rangle_{b_{6}}|H\rangle_{c_{6}}\nonumber\\
&+&|V\rangle_{a_{5}}|V\rangle_{b_{5}}|V\rangle_{c_{5}}+|H\rangle_{a_{4}}|H\rangle_{b_{4}}|H\rangle_{c_{4}}).\label{evolve1}
\end{eqnarray}
From Fig. 1, items $|V\rangle_{a_{3}}|V\rangle_{b_{3}}|V\rangle_{c_{3}}$ and $|H\rangle_{a_{4}}|H\rangle_{b_{4}}|H\rangle_{c_{4}}$ will couple in the BD$_{1}$, BD$_{2}$, and BD$_{3}$, respectively, and finally become the polarization entangled state $|\Phi_{0}^{+}\rangle$ in the output modes D$_{1}$D$_{2}$D$_{3}$. On the other hand, items $|H\rangle_{a_{6}}|H\rangle_{b_{6}}|H\rangle_{c_{6}}$ and $|V\rangle_{a_{5}}|V\rangle_{b_{5}}|V\rangle_{c_{5}}$ will also couple in the BD$_{4}$, BD$_{5}$ and BD$_{6}$ and become the polarization entangled state $|\Phi_{0}^{+}\rangle$ in the output modes D$_{4}$D$_{5}$D$_{6}$.

The second case $|\Phi_{1}^{+}\rangle_{ABC}\otimes|\phi_{1}^{+}\rangle_{ABC}$ can evolve as
\begin{eqnarray}
&&|\Phi_{1}^{+}\rangle_{ABC}\otimes|\phi_{1}^{+}\rangle_{ABC}\nonumber\\
&=&\frac{1}{\sqrt{2}}(|H\rangle_{A}|H\rangle_{B}|V\rangle_{C}+|V\rangle_{A}|V\rangle_{B}|H\rangle_{C})\nonumber\\
&\otimes&\frac{1}{\sqrt{2}}(|a_{1}\rangle_{A}|b_{1}\rangle_{B}|c_{2}\rangle_{C}+|a_{2}\rangle_{A}|b_{2}\rangle_{B}|c_{1}\rangle_{C})\nonumber\\
&=&\frac{1}{2}(|H\rangle_{a_{1}}|H\rangle_{b_{1}}|V\rangle_{c_{2}}+|H\rangle_{a_{2}}|H\rangle_{b_{2}}|V\rangle_{c_{1}}\nonumber\\
&+&|V\rangle_{a_{1}}|V\rangle_{b_{1}}|H\rangle_{c_{2}}+|V\rangle_{a_{2}}|V\rangle_{b_{2}}|H\rangle_{c_{1}})\nonumber\\
&\rightarrow&\frac{1}{2}(|V\rangle_{a_{3}}|V\rangle_{b_{3}}|H\rangle_{c_{4}}+|H\rangle_{a_{6}}|H\rangle_{b_{6}}|V\rangle_{c_{5}}\nonumber\\
&+&|V\rangle_{a_{5}}|V\rangle_{b_{5}}|H\rangle_{c_{6}}+|H\rangle_{a_{4}}|H\rangle_{b_{4}}|V\rangle_{c_{3}}).\label{evolve1}
\end{eqnarray}
Items $|V\rangle_{a_{3}}|V\rangle_{b_{3}}|H\rangle_{c_{4}}$ and $|H\rangle_{a_{4}}|H\rangle_{b_{4}}|V\rangle_{c_{3}}$ will become $|\Phi_{1}^{+}\rangle$ in the output modes D$_{1}$D$_{2}$D$_{3}$. Items $|H\rangle_{a_{6}}|H\rangle_{b_{6}}|V\rangle_{c_{5}}$ and $|V\rangle_{a_{5}}|V\rangle_{b_{5}}|H\rangle_{c_{6}}$ will also become $|\Phi_{1}^{+}\rangle$ in the output modes D$_{4}$D$_{5}$D$_{6}$.

On the other hand, the cases $|\Phi_{0}^{+}\rangle_{ABC}\otimes|\phi_{1}^{+}\rangle_{ABC}$ and $|\Phi_{1}^{+}\rangle_{ABC}\otimes|\phi_{0}^{+}\rangle_{ABC}$ cannot make the three photons in the output modes D$_{1}$D$_{2}$D$_{3}$ or D$_{4}$D$_{5}$D$_{6}$. For example, $|\Phi_{0}^{+}\rangle_{ABC}\otimes|\phi_{1}^{+}\rangle_{ABC}$ will lead the three photons become $|\Phi_{1}^{+}\rangle$ in output modes D$_{1}$D$_{2}$D$_{6}$ or D$_{4}$D$_{5}$D$_{3}$.  Case $|\Phi_{1}^{+}\rangle_{ABC}\otimes|\phi_{0}^{+}\rangle_{ABC}$ will become $|\Phi_{0}^{+}\rangle$ in the output modes  D$_{1}$D$_{2}$D$_{6}$ or D$_{4}$D$_{5}$D$_{3}$. In this way, by selecting the output modes D$_{1}$D$_{2}$D$_{3}$ or D$_{4}$D$_{5}$D$_{6}$ each having a photon, they can ultimately obtain a new mixed state
\begin{eqnarray}
\rho'_{ABC}=F'|\Phi_{0}^{+}\rangle_{ABC}\langle\Phi_{0}^{+}|+(1-F')|\Phi_{1}^{+}\rangle_{ABC}\langle\Phi_{1}^{+}|.
\end{eqnarray}
Here $F'$ is
\begin{eqnarray}
F'=\frac{F_{1}F_{2}}{F_{1}F_{2}+(1-F_{1})(1-F_{2})}.\label{mixed2}
\end{eqnarray}
Obviously, if $F_{1}>\frac{1}{2}$ and $F_{2}>\frac{1}{2}$, we can obtain $F'>F_{1}$ and $F'>F_{2}$. In this way, we complete the purification of bit-flip error and the success probability is $F_{1}F_{2}+(1-F_{1})(1-F_{2})$.

\section{Multipartite entanglement purification for phase-flip error}
In this section, we will describe the purification of phase-flip error. If a phase-flip error occurs in polarization part and spatial mode part,
the mixed state can be written as
\begin{eqnarray}
\varrho_{ABC}=\varrho^{P}_{ABC}\otimes\varrho^{S}_{ABC}.\label{mixed3}
\end{eqnarray}
Here
\begin{eqnarray}
\varrho^{P}_{ABC}=F_{3}|\Phi_{0}^{+}\rangle_{ABC}\langle\Phi_{0}^{+}|+(1-F_{3})|\Phi_{0}^{-}\rangle_{ABC}\langle\Phi_{0}^{-}|,
\end{eqnarray}
and
\begin{eqnarray}
\varrho^{S}_{ABC}=F_{4}|\phi_{0}^{+}\rangle_{ABC}\langle\phi_{0}^{+}|+(1-F_{4})|\phi_{0}^{-}\rangle_{ABC}\langle\phi_{0}^{-}|.
\end{eqnarray}
\begin{figure}[!h]%[tpb]
\begin{center}
\includegraphics[width=9cm,angle=0]{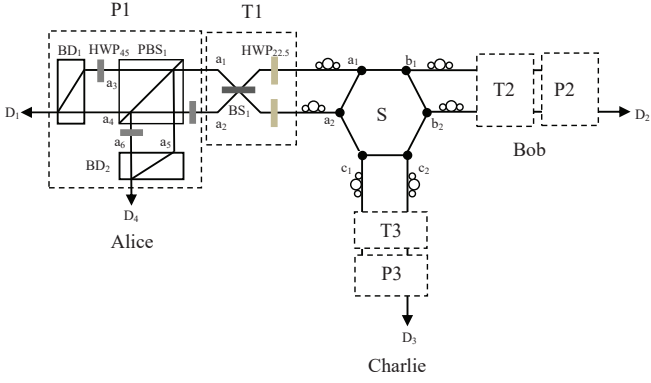}
\caption{Schematic drawing showing the principle of entanglement purification for phase-flip error. The HWP$_{22.5}$ can transform the $|H\rangle$ polarization to $\frac{1}{\sqrt{2}}(|H\rangle+|V\rangle)$ and $|V\rangle$ polarization to $\frac{1}{\sqrt{2}}(|H\rangle-|V\rangle)$. The BS is the 50:50 beam splitter. $|a_{1}\rangle\rightarrow\frac{1}{\sqrt{2}}(|a_{1}\rangle+|a_{2}\rangle)$ and $|a_{2}\rangle\rightarrow\frac{1}{\sqrt{2}}(|a_{1}\rangle-|a_{2}\rangle)$. Therefore, the  HWP$_{22.5}$ and BS both act as the role of Hadamard operation for polarization and spatial mode qubits, respectively.}
\end{center}
\end{figure}
The principle of phase-flip error is shown in Fig.2. Before purification, they should transform the phase-flip error to bit-flip error using the setup $T_{i}, i=1, 2, 3$. Here HWP$_{22.5}$ can perform the Hadamard operation and  make $|H\rangle\rightarrow\frac{1}{\sqrt{2}}(|H\rangle+|V\rangle)$ and $|V\rangle\rightarrow\frac{1}{\sqrt{2}}(|H\rangle-|V\rangle)$. The beam splitter (BS) can also act as the role of Hadamard operation for spatial mode qubit. It can make $|a_{1}\rangle\rightarrow\frac{1}{\sqrt{2}}(|a_{1}\rangle+|a_{2}\rangle)$ and $|a_{2}\rangle\rightarrow\frac{1}{\sqrt{2}}(|a_{1}\rangle-|a_{2}\rangle)$. After performing the Hadamard operation, the GHZ states in polarization and spatial mode as shown in Eq. (\ref{polarization}) and (\ref{spatial}) can be rewritten as
\begin{eqnarray}
|\Psi_{0}^{+}\rangle_{ABC}&=&\frac{1}{2}(|H\rangle_{A}|H\rangle_{B}|H\rangle_{C}+|H\rangle_{A}|V\rangle_{B}|V\rangle_{C}\nonumber\\
&+&|V\rangle_{A}|H\rangle_{B}|V\rangle_{C}+|V\rangle_{A}|V\rangle_{B}|H\rangle_{C}),\nonumber\\
|\Psi_{0}^{-}\rangle_{ABC}&=&\frac{1}{2}(|H\rangle_{A}|H\rangle_{B}|V\rangle_{C}+|H\rangle_{A}|V\rangle_{B}|H\rangle_{C}\nonumber\\
&+&|V\rangle_{A}|H\rangle_{B}|H\rangle_{C}+|V\rangle_{A}|V\rangle_{B}|V\rangle_{C}),\nonumber\\
|\Psi_{1}^{+}\rangle_{ABC}&=&\frac{1}{2}(|H\rangle_{A}|H\rangle_{B}|H\rangle_{C}+|H\rangle_{A}|V\rangle_{B}|V\rangle_{C}\nonumber\\
&-&|V\rangle_{A}|H\rangle_{B}|V\rangle_{C}-|V\rangle_{A}|V\rangle_{B}|H\rangle_{C}),\nonumber\\
|\Psi_{1}^{-}\rangle_{ABC}&=&\frac{1}{2}(|H\rangle_{A}|H\rangle_{B}|V\rangle_{C}+|H\rangle_{A}|V\rangle_{B}|H\rangle_{C}\nonumber\\
&-&|V\rangle_{A}|H\rangle_{B}|H\rangle_{C}-|V\rangle_{A}|V\rangle_{B}|V\rangle_{C}),\nonumber\\
|\Psi_{2}^{+}\rangle_{ABC}&=&\frac{1}{2}(|H\rangle_{A}|H\rangle_{B}|H\rangle_{C}-|H\rangle_{A}|V\rangle_{B}|V\rangle_{C}\nonumber\\
&+&|V\rangle_{A}|H\rangle_{B}|V\rangle_{C}-|V\rangle_{A}|V\rangle_{B}|H\rangle_{C}),\nonumber\\
|\Psi_{2}^{-}\rangle_{ABC}&=&\frac{1}{2}(|H\rangle_{A}|H\rangle_{B}|V\rangle_{C}-|H\rangle_{A}|V\rangle_{B}|H\rangle_{C}\nonumber\\
&+&|V\rangle_{A}|H\rangle_{B}|H\rangle_{C}-|V\rangle_{A}|V\rangle_{B}|V\rangle_{C}),\nonumber\\
|\Psi_{3}^{+}\rangle_{ABC}&=&\frac{1}{2}(|H\rangle_{A}|H\rangle_{B}|H\rangle_{C}-|H\rangle_{A}|V\rangle_{B}|V\rangle_{C}\nonumber\\
&-&|V\rangle_{A}|H\rangle_{B}|V\rangle_{C}+|V\rangle_{A}|V\rangle_{B}|H\rangle_{C}),\nonumber\\
|\Psi_{3}^{-}\rangle_{ABC}&=&\frac{1}{2}(|H\rangle_{A}|H\rangle_{B}|V\rangle_{C}-|H\rangle_{A}|V\rangle_{B}|H\rangle_{C}\nonumber\\
&-&|V\rangle_{A}|H\rangle_{B}|H\rangle_{C}+|V\rangle_{A}|V\rangle_{B}|V\rangle_{C}),\nonumber\\
\end{eqnarray}
\begin{eqnarray}
|\psi_{0}^{+}\rangle_{ABC}&=&\frac{1}{2}(|a_{1}\rangle_{A}|b_{1}\rangle_{B}|c_{1}\rangle_{C}+|a_{1}\rangle_{A}|b_{2}\rangle_{B}|c_{2}\rangle_{C}\nonumber\\
&+&|a_{2}\rangle_{A}|b_{1}\rangle_{B}|c_{2}\rangle_{C}+|a_{2}\rangle_{A}|b_{2}\rangle_{B}|c_{1}\rangle_{C}),\nonumber\\
|\psi_{0}^{-}\rangle_{ABC}&=&\frac{1}{2}(|a_{1}\rangle_{A}|b_{1}\rangle_{B}|c_{2}\rangle_{C}+|a_{1}\rangle_{A}|b_{2}\rangle_{B}|c_{1}\rangle_{C}\nonumber\\
&+&|a_{2}\rangle_{A}|b_{1}\rangle_{B}|c_{1}\rangle_{C}+|a_{2}\rangle_{A}|b_{2}\rangle_{B}|c_{2}\rangle_{C}),\nonumber\\
|\psi_{1}^{+}\rangle_{ABC}&=&\frac{1}{2}(|a_{1}\rangle_{A}|b_{1}\rangle_{B}|c_{1}\rangle_{C}+|a_{1}\rangle_{A}|b_{2}\rangle_{B}|c_{2}\rangle_{C}\nonumber\\
&-&|a_{2}\rangle_{A}|b_{1}\rangle_{B}|c_{2}\rangle_{C}-|a_{2}\rangle_{A}|b_{2}\rangle_{B}|c_{1}\rangle_{C}),\nonumber\\
|\psi_{1}^{-}\rangle_{ABC}&=&\frac{1}{2}(|a_{1}\rangle_{A}|b_{1}\rangle_{B}|c_{2}\rangle_{C}+|a_{1}\rangle_{A}|b_{2}\rangle_{B}|c_{1}\rangle_{C}\nonumber\\
&-&|a_{2}\rangle_{A}|b_{1}\rangle_{B}|c_{1}\rangle_{C}-|a_{2}\rangle_{A}|b_{2}\rangle_{B}|c_{2}\rangle_{C}),\nonumber\\
|\psi_{2}^{+}\rangle_{ABC}&=&\frac{1}{2}(|a_{1}\rangle_{A}|b_{1}\rangle_{B}|c_{1}\rangle_{C}-|a_{1}\rangle_{A}|b_{2}\rangle_{B}|c_{2}\rangle_{C}\nonumber\\
&+&|a_{2}\rangle_{A}|b_{1}\rangle_{B}|c_{2}\rangle_{C}-|a_{2}\rangle_{A}|b_{2}\rangle_{B}|c_{1}\rangle_{C}),\nonumber\\
|\psi_{2}^{-}\rangle_{ABC}&=&\frac{1}{2}(|a_{1}\rangle_{A}|b_{1}\rangle_{B}|c_{2}\rangle_{C}-|a_{1}\rangle_{A}|b_{2}\rangle_{B}|c_{1}\rangle_{C}\nonumber\\
&+&|a_{2}\rangle_{A}|b_{1}\rangle_{B}|c_{1}\rangle_{C}-|a_{2}\rangle_{A}|b_{2}\rangle_{B}|c_{2}\rangle_{C}),\nonumber\\
|\psi_{3}^{+}\rangle_{ABC}&=&\frac{1}{2}(|a_{1}\rangle_{A}|b_{1}\rangle_{B}|c_{1}\rangle_{C}-|a_{1}\rangle_{A}|b_{2}\rangle_{B}|c_{2}\rangle_{C}\nonumber\\
&-&|a_{2}\rangle_{A}|b_{1}\rangle_{B}|c_{2}\rangle_{C}+|a_{2}\rangle_{A}|b_{2}\rangle_{B}|c_{1}\rangle_{C}),\nonumber\\
|\psi_{3}^{-}\rangle_{ABC}&=&\frac{1}{2}(|a_{1}\rangle_{A}|b_{1}\rangle_{B}|c_{2}\rangle_{C}-|a_{1}\rangle_{A}|b_{2}\rangle_{B}|c_{1}\rangle_{C}\nonumber\\
&-&|a_{2}\rangle_{A}|b_{1}\rangle_{B}|c_{1}\rangle_{C}+|a_{2}\rangle_{A}|b_{2}\rangle_{B}|c_{2}\rangle_{C}).\nonumber\\
\end{eqnarray}
After performing the Hadamard operations, the mixed state in Eq.(\ref{mixed3}) can be rewritten as
\begin{eqnarray}
\sigma_{ABC}=\sigma^{P}_{ABC}\otimes\sigma^{S}_{ABC}.\label{mixed4}
\end{eqnarray}
Here
\begin{eqnarray}
\sigma^{P}_{ABC}=F_{3}|\Psi_{0}^{+}\rangle_{ABC}\langle\Psi_{0}^{+}|+(1-F_{3})|\Psi_{0}^{-}\rangle_{ABC}\langle\Psi_{0}^{-}|,
\end{eqnarray}
and
\begin{eqnarray}
\sigma^{S}_{ABC}=F_{4}|\psi_{0}^{+}\rangle_{ABC}\langle\psi_{0}^{+}|+(1-F_{4})|\psi_{0}^{-}\rangle_{ABC}\langle\psi_{0}^{-}|.
\end{eqnarray}
Therefore, $\sigma_{ABC}$ can be described as follows. With the probability of $F_{3}F_{4}$, it is in the state $|\Psi_{0}^{+}\rangle_{ABC}\otimes|\psi_{0}^{+}\rangle_{ABC}$. With the probability of $F_{3}(1-F_{4})$, it is in the state $|\Psi_{0}^{+}\rangle_{ABC}\otimes|\psi_{0}^{-}\rangle_{ABC}$. With the probability of $(1-F_{3})F_{4}$, it is in the state $|\Psi_{0}^{-}\rangle_{ABC}\otimes|\psi_{0}^{+}\rangle_{ABC}$. With the probability of $(1-F_{3})(1-F_{4})$, it is in the state $|\Psi_{0}^{-}\rangle_{ABC}\otimes|\psi_{0}^{-}\rangle_{ABC}$. We first discuss the case $|\Psi_{0}^{+}\rangle_{ABC}\otimes|\psi_{0}^{+}\rangle_{ABC}$. It will evolve as
\begin{eqnarray}
&&|\Psi_{0}^{+}\rangle_{ABC}\otimes|\psi_{0}^{+}\rangle_{ABC}\nonumber\\
&=&\frac{1}{2}(|H\rangle_{A}|H\rangle_{B}|H\rangle_{C}+|H\rangle_{A}|V\rangle_{B}|V\rangle_{C}\nonumber\\
&+&|V\rangle_{A}|H\rangle_{B}|V\rangle_{C}+|V\rangle_{A}|V\rangle_{B}|H\rangle_{C})\nonumber\\
&\otimes&\frac{1}{2}(|a_{1}\rangle_{A}|b_{1}\rangle_{B}|c_{1}\rangle_{C}+|a_{1}\rangle_{A}|b_{2}\rangle_{B}|c_{2}\rangle_{C}\nonumber\\
&+&|a_{2}\rangle_{A}|b_{1}\rangle_{B}|c_{2}\rangle_{C}+|a_{2}\rangle_{A}|b_{2}\rangle_{B}|c_{1}\rangle_{C})\nonumber\\
&=&\frac{1}{4}(|H\rangle_{a_{1}}|H\rangle_{b_{1}}|H\rangle_{c_{1}}+|H\rangle_{a_{1}}|H\rangle_{b_{2}}|H\rangle_{c_{2}}\nonumber\\
&+&|H\rangle_{a_{2}}|H\rangle_{b_{1}}|H\rangle_{c_{2}}+|H\rangle_{a_{2}}|H\rangle_{b_{2}}|H\rangle_{c_{1}}\nonumber\\
&+&|H\rangle_{a_{1}}|V\rangle_{b_{1}}|V\rangle_{c_{1}}+|H\rangle_{a_{1}}|V\rangle_{b_{2}}|V\rangle_{c_{2}}\nonumber\\
&+&|H\rangle_{a_{2}}|V\rangle_{b_{1}}|V\rangle_{c_{2}}+|H\rangle_{a_{2}}|V\rangle_{b_{2}}|V\rangle_{c_{1}}\nonumber\\
&+&|V\rangle_{a_{1}}|H\rangle_{b_{1}}|V\rangle_{c_{1}}+|V\rangle_{a_{1}}|H\rangle_{b_{2}}|V\rangle_{c_{2}}\nonumber\\
&+&|V\rangle_{a_{2}}|H\rangle_{b_{1}}|V\rangle_{c_{2}}+|V\rangle_{a_{2}}|H\rangle_{b_{2}}|V\rangle_{c_{1}}\nonumber\\
&+&|V\rangle_{a_{1}}|V\rangle_{b_{1}}|H\rangle_{c_{1}}+|V\rangle_{a_{1}}|V\rangle_{b_{2}}|H\rangle_{c_{2}}\nonumber\\
&+&|V\rangle_{a_{2}}|V\rangle_{b_{1}}|H\rangle_{c_{2}}+|V\rangle_{a_{2}}|V\rangle_{b_{2}}|H\rangle_{c_{1}})\nonumber\\
&\rightarrow&\frac{1}{4}(|V\rangle_{a_{3}}|V\rangle_{b_{3}}|V\rangle_{c_{3}}+|V\rangle_{a_{3}}|H\rangle_{b_{6}}|H\rangle_{c_{6}}\nonumber\\
&+&|H\rangle_{a_{6}}|V\rangle_{b_{3}}|H\rangle_{c_{6}}+|H\rangle_{a_{6}}|H\rangle_{b_{6}}|V\rangle_{c_{3}}\nonumber\\
&+&|V\rangle_{a_{3}}|V\rangle_{b_{5}}|V\rangle_{c_{5}}+|V\rangle_{a_{3}}|H\rangle_{b_{4}}|H\rangle_{c_{4}}\nonumber\\
&+&|H\rangle_{a_{6}}|V\rangle_{b_{5}}|H\rangle_{c_{4}}+|H\rangle_{a_{6}}|H\rangle_{b_{4}}|V\rangle_{c_{5}}\nonumber\\
&+&|V\rangle_{a_{5}}|V\rangle_{b_{3}}|V\rangle_{c_{5}}+|V\rangle_{a_{5}}|H\rangle_{b_{6}}|H\rangle_{c_{4}}\nonumber\\
&+&|H\rangle_{a_{4}}|V\rangle_{b_{3}}|H\rangle_{c_{4}}+|H\rangle_{a_{4}}|H\rangle_{b_{6}}|V\rangle_{c_{5}}\nonumber\\
&+&|V\rangle_{a_{5}}|V\rangle_{b_{5}}|V\rangle_{c_{3}}+|V\rangle_{a_{5}}|H\rangle_{b_{4}}|H\rangle_{c_{6}}\nonumber\\
&+&|H\rangle_{a_{4}}|V\rangle_{b_{5}}|H\rangle_{c_{6}}+|H\rangle_{a_{4}}|H\rangle_{b_{4}}|V\rangle_{c_{3}}).\label{evolve2}
\end{eqnarray}
From Eq. (\ref{evolve2}), item $|V\rangle_{a_{3}}|V\rangle_{b_{3}}|V\rangle_{c_{3}}$, $|V\rangle_{a_{3}}|H\rangle_{b_{4}}|H\rangle_{c_{4}}$, $|H\rangle_{a_{4}}|V\rangle_{b_{3}}|H\rangle_{c_{4}}$ and $|H\rangle_{a_{4}}|H\rangle_{b_{4}}|V\rangle_{c_{3}}$ will be in the output modes D$_{1}$D$_{2}$D$_{3}$ and become
\begin{eqnarray}
&&\frac{1}{2}(|V\rangle_{D_{1}}|V\rangle_{D_{2}}|V\rangle_{D_{3}}+|V\rangle_{D_{1}}|H\rangle_{D_{2}}|H\rangle_{D_{3}}\nonumber\\
&+&|H\rangle_{D_{1}}|V\rangle_{D_{2}}|H\rangle_{D_{3}}+|H\rangle_{D_{1}}|H\rangle_{D_{2}}|V\rangle_{D_{3}}).\label{evolve3}
\end{eqnarray}
By performing bit-flip operation on each photon, state in Eq.(\ref{evolve3}) can be changed to $|\Psi_{0}^{+}\rangle_{ABC}$. Finally, they can change $|\Psi_{0}^{+}\rangle_{ABC}$ to $|\Phi_{0}^{+}\rangle_{ABC}$ by adding another  Hadamard operation on each photon. On the other hand, from Eq.(\ref{evolve2}), by selecting the output modes D$_{1}$D$_{5}$D$_{6}$, D$_{4}$D$_{5}$D$_{3}$, or D$_{4}$D$_{2}$D$_{6}$, they can obtain the same state in Eq.(\ref{evolve3}). In this way, they can also obtain  $|\Phi_{0}^{+}\rangle_{ABC}$.

The case $|\Psi_{0}^{-}\rangle_{ABC}\otimes|\psi_{0}^{-}\rangle_{ABC}$ can be evolve as
\begin{eqnarray}
&&|\Psi_{0}^{-}\rangle_{ABC}\otimes|\psi_{0}^{-}\rangle_{ABC}\nonumber\\
&=&\frac{1}{2}(|H\rangle_{A}|H\rangle_{B}|V\rangle_{C}+|H\rangle_{A}|V\rangle_{B}|H\rangle_{C}\nonumber\\
&+&|V\rangle_{A}|H\rangle_{B}|H\rangle_{C}+|V\rangle_{A}|V\rangle_{B}|V\rangle_{C})\nonumber\\
&&\otimes\frac{1}{2}(|a_{1}\rangle_{A}|b_{1}\rangle_{B}|c_{2}\rangle_{C}+|a_{1}\rangle_{A}|b_{2}\rangle_{B}|c_{1}\rangle_{C}\nonumber\\
&+&|a_{2}\rangle_{A}|b_{1}\rangle_{B}|c_{1}\rangle_{C}+|a_{2}\rangle_{A}|b_{2}\rangle_{B}|c_{2}\rangle_{C})\nonumber\\
&&\rightarrow\frac{1}{4}(|V\rangle_{a_{3}}|V\rangle_{b_{3}}|H\rangle_{c_{4}}+|V\rangle_{a_{3}}|H\rangle_{b_{6}}|V\rangle_{c_{5}}\nonumber\\
&+&|H\rangle_{a_{6}}|V\rangle_{b_{3}}|V\rangle_{c_{5}}+|H\rangle_{a_{6}}|H\rangle_{b_{6}}|H\rangle_{c_{4}}\nonumber\\
&+&|V\rangle_{a_{3}}|V\rangle_{b_{5}}|H\rangle_{c_{6}}+|V\rangle_{a_{3}}|H\rangle_{b_{4}}|V\rangle_{c_{3}}\nonumber\\
&+&|H\rangle_{a_{6}}|V\rangle_{b_{5}}|V\rangle_{c_{3}}+|H\rangle_{a_{6}}|H\rangle_{b_{6}}|H\rangle_{c_{6}}\nonumber\\
&+&|V\rangle_{a_{5}}|V\rangle_{b_{3}}|H\rangle_{c_{6}}+|V\rangle_{a_{5}}|H\rangle_{b_{6}}|V\rangle_{c_{3}}\nonumber\\
&+&|H\rangle_{a_{4}}|V\rangle_{b_{3}}|V\rangle_{c_{3}}+|H\rangle_{a_{4}}|H\rangle_{b_{6}}|H\rangle_{c_{6}}\nonumber\\
&+&|V\rangle_{a_{5}}|V\rangle_{b_{5}}|H\rangle_{c_{4}}+|V\rangle_{a_{5}}|H\rangle_{b_{4}}|H\rangle_{c_{5}}\nonumber\\
&+&|H\rangle_{a_{4}}|V\rangle_{b_{5}}|V\rangle_{c_{5}}+|H\rangle_{a_{4}}|H\rangle_{b_{4}}|H\rangle_{c_{4}}).\label{evolve4}
\end{eqnarray}
From Eq.(\ref{evolve4}), items $|V\rangle_{a_{3}}|V\rangle_{b_{3}}|H\rangle_{c_{4}}$, $|V\rangle_{a_{3}}|H\rangle_{b_{4}}|V\rangle_{c_{3}}$,
$|H\rangle_{a_{4}}|V\rangle_{b_{3}}|V\rangle_{c_{3}}$ and $|H\rangle_{a_{4}}|H\rangle_{b_{4}}|H\rangle_{c_{4}}$ will be in the output modes D$_{1}$D$_{2}$D$_{3}$ and become
\begin{eqnarray}
&&\frac{1}{2}(|V\rangle_{D_{1}}|V\rangle_{D_{2}}|H\rangle_{D_{3}}+|V\rangle_{D_{1}}|H\rangle_{D_{2}}|V\rangle_{D_{3}}\nonumber\\
&+&|H\rangle_{D_{1}}|V\rangle_{D_{2}}|V\rangle_{D_{3}}+|H\rangle_{D_{1}}|H\rangle_{D_{2}}|H\rangle_{D_{3}}).\label{evolve5}
\end{eqnarray}
State in Eq.(\ref{evolve5}) can be changed to $|\Psi_{0}^{-}\rangle_{ABC}$ by adding bit-flip operation on each photon. By adding another Hadamard operation on each photon, $|\Psi_{0}^{-}\rangle_{ABC}$ can be converted to $|\Phi_{0}^{-}\rangle_{ABC}$. On the other hand, from Eq.(\ref{evolve4}), by selecting the output modes D$_{1}$D$_{5}$D$_{6}$, D$_{4}$D$_{5}$D$_{3}$, or D$_{4}$D$_{2}$D$_{6}$, they can obtain the same state in Eq.(\ref{evolve5}). In this way, they can also obtain  $|\Phi_{0}^{-}\rangle_{ABC}$.

The other cases $|\Psi_{0}^{+}\rangle_{ABC}\otimes|\psi_{0}^{-}\rangle_{ABC}$ and $|\Psi_{0}^{-}\rangle_{ABC}\otimes|\psi_{0}^{-}\rangle_{ABC}$
will lead the photons in the output modes D$_{1}$D$_{2}$D$_{6}$, D$_{1}$D$_{5}$D$_{3}$, D$_{4}$D$_{2}$D$_{3}$, or D$_{4}$D$_{5}$D$_{6}$. Therefore, by selecting the cases that  the output modes D$_{1}$D$_{2}$D$_{3}$, D$_{1}$D$_{5}$D$_{6}$, D$_{4}$D$_{5}$D$_{3}$, or D$_{4}$D$_{2}$D$_{6}$ contain one photon, cases  $|\Psi_{0}^{+}\rangle_{ABC}\otimes|\psi_{0}^{-}\rangle_{ABC}$ and $|\Psi_{0}^{-}\rangle_{ABC}\otimes|\psi_{0}^{-}\rangle_{ABC}$  can be eliminated automatically. Finally, with the probability of $F_{3}F_{4}$, they will obtain $|\Phi_{0}^{+}\rangle_{ABC}$. With the probability of $(1-F_{3})(1-F_{4})$, they will obtain $|\Phi_{0}^{-}\rangle_{ABC}$. The new mixed state can be rewritten as
\begin{eqnarray}
\sigma'_{ABC}=F''|\Psi_{0}^{+}\rangle_{ABC}\langle\Psi_{0}^{+}|+(1-F'')|\Psi_{0}^{-}\rangle_{ABC}\langle\Psi_{0}^{-}|.\nonumber\\
\end{eqnarray}
Here $F''$ is
\begin{eqnarray}
F''=\frac{F_{3}F_{4}}{F_{3}F_{4}+(1-F_{3})(1-F_{4})}.\label{mixed5}
\end{eqnarray}
Similar to Eq.(\ref{mixed2}), $F''>F_{3}$ and $F''>F_{4}$ if $F_{3}>\frac{1}{2}$ and $F_{4}>\frac{1}{2}$.

\section{Arbitrary multipartite entanglement purification}
It is easy to extend the EPP to the arbitrary GHZ state. The $m$-photon GHZ state in polarization can be described as
\begin{eqnarray}
|\Phi_{0}^{\pm}\rangle_{m}&=&\frac{1}{\sqrt{2}}(|H\rangle_{1}|H\rangle_{2}\cdots|H\rangle_{m}\pm|V\rangle_{1}|V\rangle_{2}\cdots|V\rangle_{m}),\nonumber\\
|\Phi_{1}^{\pm}\rangle_{m}&=&\frac{1}{\sqrt{2}}(|H\rangle_{1}|H\rangle_{2}\cdots|V\rangle_{m}\pm|V\rangle_{1}|V\rangle_{2}\cdots|H\rangle_{m}),\nonumber\\
&\cdots&,\nonumber\\
|\Phi_{2^{m-1}}^{\pm}\rangle_{m}&=&\frac{1}{\sqrt{2}}(|V\rangle_{1}|H\rangle_{2}\cdots|H\rangle_{m}\pm|H\rangle_{1}|V\rangle_{2}\cdots|V\rangle_{m}),\nonumber\\
\end{eqnarray}
\begin{figure}[!h]%[tpb]
\begin{center}
\includegraphics[width=8cm,angle=0]{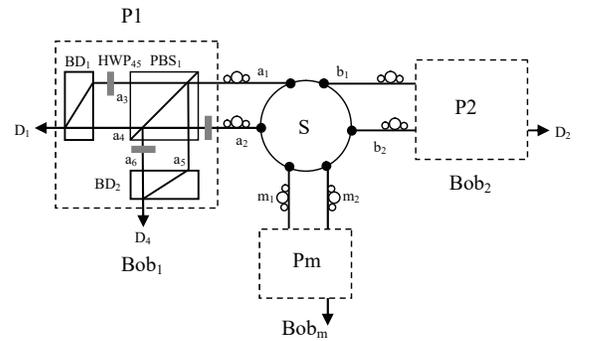}
\caption{Schematic drawing showing the principle of entanglement purification for arbitrary GHZ state.}
\end{center}
\end{figure}
On the other hand, the $m$-photon GHZ state in spatial mode can be described as
\begin{eqnarray}
|\phi_{0}^{\pm}\rangle_{m}&=&\frac{1}{\sqrt{2}}(|a_{1}\rangle_{1}|b_{1}\rangle_{2}\nonumber\\
&\cdots&|m_{1}\rangle_{m}\pm|a_{2}\rangle_{1}|b_{2}\rangle_{2}\cdots|m_{2}\rangle_{m}),\nonumber\\
|\phi_{1}^{\pm}\rangle_{m}&=&\frac{1}{\sqrt{2}}(|a_{1}\rangle_{1}|b_{1}\rangle_{2}\nonumber\\
&\cdots&|m_{2}\rangle_{m}\pm|a_{2}\rangle_{1}|b_{2}\rangle_{2}\cdots|m_{1}\rangle_{m}),\nonumber\\
&\cdots&,\nonumber\\
|\phi_{2^{m-1}}^{\pm}\rangle_{m}&=&\frac{1}{\sqrt{2}}(|a_{2}\rangle_{1}|b_{1}\rangle_{2}\nonumber\\
&\cdots&|m_{1}\rangle_{m}\pm|a_{1}\rangle_{1}|b_{1}\rangle_{2}\cdots|m_{2}\rangle_{m}).
\end{eqnarray}
As shown in Fig. 3, the entanglement source prepares the $m$-photon hyperentangled state $|\Psi\rangle_{m}$ of the form
\begin{eqnarray}
|\Psi\rangle_{m}=|\Phi_{0}^{+}\rangle_{m}\otimes|\phi_{0}^{+}\rangle_{m}.\label{initial}
\end{eqnarray}
Such hyperentangled state is distributed to $m$ parties, named Bob$_{1}$, Bob$_{2}$, $\cdots$, and Bob$_{m}$. After distribution, the initial state becomes a mixed state as
\begin{eqnarray}
\rho_{m}=\rho^{P}_{m}\otimes\rho^{P}_{m}.\label{mixed6}
\end{eqnarray}
Here $\rho^{P}_{m}$ can be written as
\begin{eqnarray}
\rho^{P}_{m}=F_{1}|\Phi_{0}^{+}\rangle_{m}\langle\Phi_{0}^{+}|+(1-F_{_{1}})|\Phi_{1}^{+}\rangle_{m}\langle\Phi_{1}^{+}|.
\end{eqnarray}
$\rho^{P}_{m}$ can be written as
\begin{eqnarray}
\rho^{S}_{m}=F_{2}|\phi_{0}^{+}\rangle_{m}\langle\phi_{0}^{+}|+(1-F_{_{2}})|\phi_{1}^{+}\rangle_{m}\langle\phi_{1}^{+}|.
\end{eqnarray}
The purification is similar as described in above section. By selecting the output modes D$_{1}$,D$_{2}$, $\cdots$, D$_{m}$ exactly contain one photon, they can ultimately obtain a high fidelity  mixed state in polarization. The fidelity $F'$ is same as it is shown in Eq. (\ref{mixed2}). On the other hand, if the phase-flip error occurs, one can also convert it to the bit-flip error, and perform the purification in a next step. In this way, one can purify the arbitrary $m$-photon GHZ state.

\section{Discussion}
So far, we have completely described this EPP. We first described the EPP for three-photon GHZ state with a bit-flip error. Then we explained the EPP with a phase-flip error. In this way, all the errors can be purified. Finally, we extend this EPP for arbitrary GHZ state and which can be purified in the same way. In above, we suppose that the bit-flip error  occurs on the first qubit. In a practical transmission, the hyperentanglement in polarization and spatial mode will suffer from different errors. For example, the bit-flip error occurs on the first qubit in polarization and which makes the polarization part become the mixed state in Eq. (\ref{polarization1}), while the bit-flip error occurs on the second qubit in spatial mode and makes the spatial part become
\begin{eqnarray}
\rho'^{S}_{ABC}=F_{3}|\phi_{0}^{+}\rangle_{ABC}\langle\phi_{0}^{+}|+(1-F_{_{3}})|\phi_{2}^{+}\rangle_{ABC}\langle\phi_{2}^{+}|.\label{spatial1}
\end{eqnarray}
Therefore, the hyperentangled mixed state can be written as
\begin{eqnarray}
\rho'_{ABC}=\rho^{P}_{ABC}\otimes\rho'^{S}_{ABC}.\label{mixed7}
\end{eqnarray}
With the probability of F$_{1}$F$_{3}$, it is in the state  $|\Phi_{0}^{+}\rangle_{ABC}\otimes|\phi_{0}^{+}\rangle_{ABC}$. Such state will make
the three photons in the output modes D$_{1}$D$_{2}$D$_{3}$ or D$_{4}$D$_{5}$D$_{6}$. With the probability of (1-F$_{1})$F$_{3}$, it is in the state
$|\Phi_{1}^{+}\rangle_{ABC}\otimes|\phi_{0}^{+}\rangle_{ABC}$. Such state will make the three photons in the output modes D$_{1}$D$_{2}$D$_{6}$ or D$_{4}$D$_{5}$D$_{3}$. With the probability of F$_{1}$(1-F$_{3})$, it is in the state $|\Phi_{0}^{+}\rangle_{ABC}\otimes|\phi_{2}^{+}\rangle_{ABC}$. Such state will make the three photons in the output modes D$_{1}$D$_{5}$D$_{3}$ or D$_{4}$D$_{2}$D$_{6}$. Finally, with the probability of (1-F$_{1}$)(1-F$_{3}$), it is in the state $|\Phi_{1}^{+}\rangle_{ABC}\otimes|\phi_{2}^{+}\rangle_{ABC}$. Such state will make the three photons in the output modes D$_{1}$D$_{5}$D$_{6}$ or D$_{4}$D$_{2}$D$_{3}$. In this way, by selecting the cases D$_{1}$D$_{2}$D$_{3}$ or D$_{4}$D$_{5}$D$_{6}$,
they  can ultimately obtain the polarization state $|\Phi_{0}^{+}\rangle_{ABC}$. Interestingly, if the mixed state is described as shown in Eq.(\ref{mixed7}), the bit-flip error can be completely purified. The second case $|\Phi_{1}^{+}\rangle_{ABC}\otimes|\phi_{0}^{+}\rangle_{ABC}$ will lead the photons in D$_{1}$D$_{2}$D$_{6}$ or D$_{4}$D$_{5}$D$_{3}$ and such state will become $|\Phi_{1}^{+}\rangle_{ABC}$. They can add a bit-flip operation on the first photon and convert it to $|\Phi_{0}^{+}\rangle_{ABC}$ deterministically. On the other hand, the second case $|\Phi_{0}^{+}\rangle_{ABC}\otimes|\phi_{2}^{+}\rangle_{ABC}$ will lead the photons in the spatial modes D$_{1}$D$_{5}$D$_{3}$ or D$_{4}$D$_{2}$D$_{6}$ and become $|\Phi_{2}^{+}\rangle_{ABC}$. They can also add the bit-flip operation on the second photon and convert it to $|\Phi_{0}^{+}\rangle_{ABC}$ deterministically. The final case  $|\Phi_{1}^{+}\rangle_{ABC}\otimes|\phi_{2}^{+}\rangle_{ABC}$ will also become $|\Phi_{1}^{+}\rangle_{ABC}$ and can be converted to $|\Phi_{0}^{+}\rangle_{ABC}$ deterministically. In this way, they can obtain the maximally pure entangled state $|\Phi_{0}^{+}\rangle_{ABC}$ with the probability of 100$\%$. For a general mixed state with bit-flip error, the polarization part and spatial-mode part can be written as
\begin{eqnarray}
\rho''^{P}_{ABC}&=&F_{1}|\Phi_{0}^{+}\rangle_{ABC}\langle\Phi_{0}^{+}|+F_{2}|\Phi_{1}^{+}\rangle_{ABC}\langle\Phi_{1}^{+}|\nonumber\\
&+&F_{3}|\Phi_{2}^{+}\rangle_{ABC}\langle\Phi_{2}^{+}|+F_{4}|\Phi_{3}^{+}\rangle_{ABC}\langle\Phi_{3}^{+}|,\label{polarization3}
\end{eqnarray}
and
\begin{eqnarray}
\rho''^{S}_{ABC}&=&F_{4}|\phi_{0}^{+}\rangle_{ABC}\langle\phi_{0}^{+}|+F_{5}|\phi_{1}^{+}\rangle_{ABC}\langle\phi_{1}^{+}|\nonumber\\
&+&F_{6}|\phi_{2}^{+}\rangle_{ABC}\langle\phi_{2}^{+}|+F_{7}|\phi_{3}^{+}\rangle_{ABC}\langle\phi_{3}^{+}|.\label{spatial3}
\end{eqnarray}
Here $F_{1}+F_{2}+F_{3}+F_{4}=1$ and $F_{5}+F_{6}+F_{7}+F_{8}=1$. Similarly,  by selecting the
output modes D$_{1}$D$_{2}$D$_{3}$ or D$_{4}$D$_{5}$D$_{6}$, they can obtain a new mixed state as
\begin{eqnarray}
\rho_{ABC}''&=&F'_{1}|\Phi_{0}^{+}\rangle_{ABC}\langle\Phi_{0}^{+}|+F'_{2}|\Phi_{1}^{+}\rangle_{ABC}\langle\Phi_{1}^{+}|\nonumber\\
&+&F'_{3}|\Phi_{2}^{+}\rangle_{ABC}\langle\Phi_{2}^{+}|+F'_{4}|\Phi_{3}^{+}\rangle_{ABC}\langle\Phi_{3}^{+}|.
\end{eqnarray}
Here
\begin{eqnarray}
F'_{1}=\frac{F_{1}F_{5}}{F_{1}F_{5}+F_{2}F_{6}+F_{3}F_{7}+F_{4}F_{8}},\nonumber\\
F'_{2}=\frac{F_{2}F_{6}}{F_{1}F_{5}+F_{2}F_{6}+F_{3}F_{7}+F_{4}F_{8}},\nonumber\\
F'_{3}=\frac{F_{3}F_{7}}{F_{1}F_{5}+F_{2}F_{6}+F_{3}F_{7}+F_{4}F_{8}},\nonumber\\
F'_{4}=\frac{F_{4}F_{8}}{F_{1}F_{5}+F_{2}F_{6}+F_{3}F_{7}+F_{4}F_{8}}.
\end{eqnarray}
If $F_{1}>\frac{1}{2}$ and $F_{5}>\frac{1}{2}$, we can also obtain $F'_{1}> F_{1}$ and $F'_{1}> F_{5}$. In this way, we can realize the general purification.

It is interesting to calculate  the purification efficiency in a practical environment. As shown in Fig. 1, the N-photon GHZ state was distributed to $N$ parties.  The transmission efficiency is $\eta_{t}=e^{-\frac{L}{L_{0}}}$. The detector efficiency is $\eta_{d}$. The $\eta_{c}$ is the probability of coupling a photon to the single-photon detector. $L_{0}$ is the attenuation
length of the channel (25 km for commercial fibre) \cite{munro}. $L$ is transmission distance. The success probability is $p_{1}=F_{1}F_{2}+(1-F_{1})(1-F_{2})$. For N-photon purification, the total purification efficiency can be calculated as
\begin{eqnarray}
P^{N}_{one}=p_{1}\eta^{N}_{t}\eta^{N}_{d}\eta^{N}_{c}.
\end{eqnarray}
In existing multipartite EPPs \cite{multipurification1,multipurification3,multipurification4,multipurification7}, they exploit two pairs of N-photon GHZ states to perform the purification. Therefore, for linear optical system, the total purification efficiency can be calculated as
\begin{eqnarray}
P^{N}_{two}=\frac{1}{4}p_{1}\eta^{2N}_{t}\eta^{2N}_{d}\eta^{2N}_{c}.
\end{eqnarray}
The ratio of $P^{N}_{one}$ and $P^{N}_{two}$ can be calculated as
\begin{eqnarray}
R=\frac{P^{N}_{one}}{P^{N}_{two}}=\frac{4}{\eta^{N}_{t}\eta^{N}_{d}\eta^{N}_{c}}=\frac{4}{(e^{-\frac{L}{L_{0}}})^{N}\eta^{N}_{d}\eta^{N}_{c}}.
\end{eqnarray}
If we let $\eta_{d}=0.9$, $\eta_{c}=0.95$ \cite{munro}.
\begin{figure}[!h]%[tpb]
\begin{center}
\includegraphics[width=6cm,angle=0]{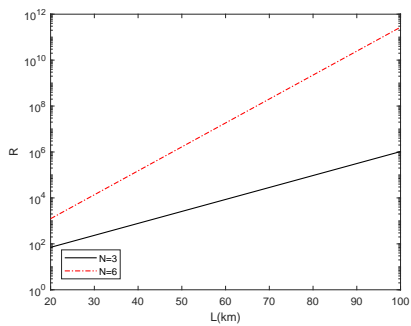}
\caption{The ratio $R$ of entanglement purification efficiency  plotted against
length of entanglement distribution. We let the photon number of GHZ state as $N=3$ and $N=6$, respectively. }
\end{center}
\end{figure}
\begin{figure}[!h]%[tpb]
\begin{center}
\includegraphics[width=6cm,angle=0]{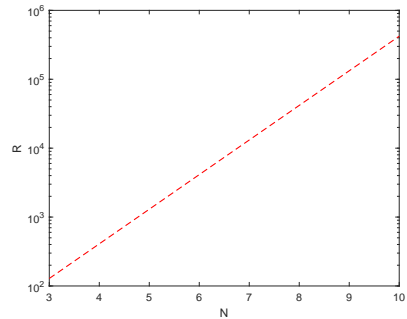}
\caption{The ratio $R$ of entanglement purification efficiency  plotted against
photon number of GHZ state. We let $L=L_{0}=25km$.}
\end{center}
\end{figure}
Fig. 4 shows the relationship between the coefficient $L$ and $R$. Here we let $N=3$ and $N=6$ respectively. We change the distance $L$ from 20km to 100km. The ratio $R$ increases rapidly. The $R$ can reach more than $10^{10}$ when $N=6$ and $L=100$km. In Fig. 5, we also calculated the $R$ altered with $N$. We let $L=L_{0}=25$km. We also showed that the $R$ increases rapidly with the photon number $N$. On the other hand, in existing multipartite EPPs \cite{multipurification1,multipurification3,multipurification4,multipurification7}, after each party performing the CNOT or similar operation, they should measure the target particles to judge that the purification is successful or not. In this EPP, the parties are not required to measure the particles and they can judge whether the purification is successful or not according to the output modes of the photons. In this way, this EPP is more economical and practical in future application.

Finally, let us briefly discuss the possible realization. This protocol mainly exploits the common linear optics, such as PBS, BS, BD, HWP. Meanwhile, this protocol require the multi-partite hyperentanglement. Such hyperentanglement was also realized in experiment \cite{source}, which show that this protocol is feasible in current experiment condition.

\section{Conclusion}
In conclusion, we have proposed the EPP for multipartite entanglement purification using hyperentanglement. After performing the EPP, the spatial entanglement can be used to purify the polarization entanglement.
Different from the previous works, this EPP has several advantages. Firstly, with the same purification success probability, this EPP only requires one pair of multipartite GHZ states, while existing EPPs usually require two pairs of multipartite GHZ state. Secondly, if consider the practical transmission and detector efficiency, this EPP may be extremely useful for the ratio of  purification efficiency increases rapidly with both the number of photons and the transmission distance. Thirdly, this protocol requires linear optics and does not add additional measurement operations, so that it is feasible for experiment.  All these advantages will make this protocol have potential application for future quantum information processing.

\section*{ACKNOWLEDGEMENTS}
This work was supported by  the National Natural Science  Foundation of China (No. 11974189).

\end{document}